 \documentclass[12pt]{article}
 \usepackage{amsfonts,amssymb}
\newtheorem{Def}{Definition}[section]

\newtheorem{The}[Def]{Theorem}

 \newcommand{\End}{\nonumber\\ }
 
 \newcommand\Mbox[1]{ \mbox{{\rm #1}}}
 \newcommand{\Textfrac}[2]{{\textstyle{\frac{#1}{#2}}}}
 
 \newcommand\Texthalf{\Textfrac{1}{2}}
 
 \newcommand{\Sum}[2]{\sum_{#1}^{#2}}
  \newcommand{\Comm}[2]{\left[ #1 , #2 \right]}
 \newcommand{\Commnd}[2]{[ #1 , #2 ]}
 \newcommand{\Pb}[2]{\left\{ #1 , #2 \right\}}
 \newcommand{\Pbnd}[2]{\{ #1 , #2 \}}
 \newcommand{\Tx}[1]{{\textstyle{{#1}}}}
 
 \newcommand{\Ssc}[1]{{\scriptscriptstyle{{#1}}}}
 \newcommand\Real{ {\mathbb R} }
 
 \newcommand{\Adjd}{ {}^{\dag}}
 
 \newcommand{\Df}{ {\rm d}}
 \newcommand{\Dbd}[1]{\frac{\Df}{\Df #1} }
 
 \newcommand{\Pbd}[1]{\Tx{\frac{\partial}{\partial #1} }}
 \newcommand{\Pbdt}[2]{\frac{\partial #1}{\partial #2}}
 \newcommand{\Dc}{ {\rm D}}
 
 \newcommand{\Curv}[4]{R_{#1 #2 #3}{}^{#4}}
  \newcommand{\Conn}[3]{\Gamma_{#1 #2}{}^{#3}}
  
 \newcommand{\Et}{\eta}
 \newcommand{\Th}{\theta }
 \newcommand{\Io}{\iota }
 
 \newcommand{\Rh}{\rho}
 
 \newcommand{\Om}{\omega}
 
 \newcommand{\Trace}{{\rm {tr}}}
 \newcommand{\Strace}{{\rm {str}}}

 \newcommand{\Man}{ \mathcal{M}}
 
 \newcommand{\Intzt}{\int_0^{t}}
  \newcommand{\Vb}{ E}
 
 \newcommand{\Ix}{ \iota_{\Ssc{X}}}
 \newcommand{\Lx}{\mathcal{L}_X }
 \newcommand{\Lxa}{\mathcal{L}_X\Adjd }
 \newcommand{\Hgm}{H_G(M) }
 \newcommand{\Gt}{\tilde{g} }
 \newcommand{\Guts}{{g^*} }
 \newcommand{\Gen}{L }
 \newcommand{\Omm}{\Omega(\Man)}
 \newcommand{\Bas}{\left( \Omm \otimes A\right)_{\rm bas} }
 \newcommand{\Wm}{W}
 \newcommand{\Basw}{\left( \Omm \otimes \Wm\right)_{\rm bas}}
 \newcommand{\Uone}{U(1)}
 \newcommand{\Dw}{\Df_{B}}
 \newcommand{\Dk}{\Df_{K}}
 \newcommand{\Dke}{\Df +  u \Pbd{\Th}   +u\, \Ix - \Th \, \Lx }
 \newcommand{\Dcc}{\Df_{C}}
 
 \newcommand{\Brst}{\textsc{brst}}
 \newcommand{\Bv}{\textsc{bv}}
 \newcommand{\Brstop}{Q}
 \newcommand{\Brstope}{Q_e}
 \newcommand{\Gff}{\chi}
 \newcommand{\Gffe}{\chi}
 \newcommand{\Hamg}{H_{g}}
 
 \newcommand{\Hamo}{H_{o}}
 
 \newcommand{\Da}{\Df\Adjd}
 \newcommand{\Fun}{{\mathcal{F}}}
 \newcommand{\Gun}{{\mathcal{G}}}
 \newcommand{\Vun}{{\mathcal{V}}}
 \newcommand{\Thun}{{\mathcal{T}}}
 
 \newcommand{\Ua}{u}
 \newcommand{\Cons}{s}
 \newcommand{\Conr}{r}
 \newcommand{\Conlp}{2}
 \newcommand{\Conl}{2}
 \newcommand{\Odd}[1]{\Pi\left( #1 \right)}
 
\begin{document}
 \begin{center}
 {\large Gauge fixing and equivariant cohomology}\\
 \ \\
   Alice Rogers\\
  \ \\
  Department of Mathematics      \\
  King's College             \\
  Strand, London  WC2R 2LS         \\
 \end{center}

 \vskip0.2in
 \begin{center}
 May 2005
 \end{center}
 \vskip0.2in
 \begin{abstract}
The supersymmetric model developed by Witten \cite{Witten82} to
study the equivariant cohomology of a manifold with an isometric
circle action is derived from the \Brst\ quantization of a simple
classical model. The gauge-fixing process is carefully analysed,
and demonstrates that different choices of gauge-fixing fermion
can lead to different quantum theories.
 \end{abstract}
 \section{Introduction}
This paper describes the quantization of a topological model whose
matrix elements give information about the equivariant cohomology
of a manifold with circle action. A key element of the
construction is the gauge-fixing fermion for the path integral
\Brst\ quantization, which takes a novel form in order to project
onto the equivariant cohomology. The model obtained by this
\Brst\ quantization process is one of two supersymmetric models
described by Witten in \cite{Witten82}; the model has been used to
derive new proofs of the character-valued index theorem
\cite{Alvare} and of various localisation formulae such as that of
\cite{DuiHec}. A mathematical account of much of this work may be found in the
books of Berline, Getzler and Vergne \cite{BerGetVer04}, of
Duistermaat \cite{Duiste96} and of Guillemin and Sternberg
\cite{GuiSte}, as well as in the references below to topological
quantum theory. Superspace path integral methods as developed in
\cite{SCSTWO,SBMS} can also be used to carry out mathematically
rigorous calculations in these models.

Topological models are of interest in both mathematics and in
physics. In physics, although they have no dynamical degrees of
freedom, topological models may correspond to a highly symmetric
phase of a theory, and give some information about phases with
broken symmetry and propagating degrees of freedom; also,
topological methods can make possible calculations outside the usual
perturbative regime such as the multi-instanton calculations of
Bruzzo et al \cite{BruFuc}. In mathematics topological quantum field
theories have led to an astonishing range of results; in general
these results have not been established by rigorous mathematical
methods, but non-rigorous functional integral techniques have
suggested a plethora of deep results which have been shown to be
true by more conventional mathematical arguments. The high degree of
symmetry of the models generally means that the functional integrals
involved descend to integrals on some finite dimensional moduli
space, but the derivation of the measure is far from trivial, and is
the nub of the issue. A recent account of the use of some
topological models to give mathematical results  may be found in
\cite{Mourao}, while reviews of topological quantum theory include
\cite{BirBla} and \cite{CorMoo}.

The astonishing power of physics to inspire these conjectures
indicates that putting the methods used in the physics literature
onto a rigorous mathematical basis is an important task. This
paper provides a step in this programme. Underlying the paper are
three beliefs: first, the canonical approach is required if the
measure on the moduli space involved is to be soundly derived;
second, although the standard form of a topological model may be
as a supersymmetric or cohomological model, sometimes referred to
as a Witten model, the more fundamental model is a classical model
with a high degree of symmetry, so high that it is evident that
the gauge invariant quantities calculated in the theory must be
topological invariants; thirdly, in topological models the
important results survive rescaling, so that it is possible to
actually perform path integral calculations rigorously by
proceeding to a limit.

In this paper a topological quantum mechanics model, referred to as
the Killing model, is described; because the model is a quantum
mechanical model the more serious analytical problems of functional
integration do not have to be tackled, but even in this simpler case
determining an appropriate gauge fixing mechanism, so that the
correct path integral measure is used, requires novel ideas. The
example in this paper demonstrates that different gauge fixing
functions can lead to different models.

The structure of this paper is that in section \ref{MODsec} the
classical Killing model is constructed, staring from a simple
Lagrangian.  Quantization is then carried out and the \Brst\
operator is constructed.  Section
\ref{ECsec} contains a brief account of the relevant background from
equivariant cohomology, including the work of Kalkman on \Brst\
operators and equivariant cohomology \cite{Kalkma}, while section
\ref{GFsec} describes the theory of gauge-fixing.  In section \ref{GFKMsec} the
\Brst\ quantization of the Killing model is completed by the construction
of the gauge-fixing fermion which underpins the whole process. This
leads to a gauge-fixed Hamiltonian which allows the rigorous
path-integral quantization of the model.

One result of this paper is to relate the two supersymmetric
models introduced by Witten in his seminal 1982 paper
\cite{Witten82} on supersymmetry and Morse theory.  The two models
seem significantly different, one concerns a Morse function  on a
manifold $\Man$ while the other considers a circle action on a
manifold. However the second can  be regarded as an example of the
first, but on a bundle over $\Man$, rather than simply on $\Man$.
In this paper the second  model is actually constructed in two
ways, first in section
\ref{MODsec} using a Morse function which is locally defined,
and the second (in the final section) using as Morse function the
moment map of the induced circle action on the (odd) cotangent
bundle $\Man$.
\section{The model}\label{MODsec}
In this section the fields and action for the topological model
studied in this paper, which will be referred to as the Killing
model, are described, together with the canonical \Brst\
quantization of the model. The basic ingredients are an
$n$-dimensional Riemannian manifold $\Man$  with
metric $g$, together with an isometric circle action on $\Man$ with
corresponding Killing vector $X$. The model obtained after
quantization is equivalent to the supersymmetric model constructed
by Witten \cite{Witten82} to study the equivariant cohomology of
$\Man$ under the circle action, as is explained at the end of
section \ref{GFKMsec}.

Before constructing this model, a brief resum\'e is given of the
Morse theory model also introduced by Witten \cite{Witten82} and
further studied in \cite{TPMT}, because the two models have many
common features. However the Killing model will lead to the
equivariant cohomology of
$\Man$, while the Morse theory model leads to the ordinary de Rham
cohomology. The gauge fixing plays a decisive role in this
differentiation, different gauge fixing fermions for the same \Brst\
model lead to quite different theories.

The action for the Morse theory model is built from a function $h$
mapping the manifold $\Man$ into the real line $\Real$, together
with a field $x:[0,t] \to \Man$, that is, a path in $\Man$. The
action is defined to be \cite{TPMT}
 \begin{equation}\label{ACTIONeq}
  S(x(.)) = \Intzt x^*( \Df h) = \Intzt \partial_i h (x(t)) \dot{x}^i(t) \,\Df t
 \end{equation}
where $\dot{x}^i$ is the time derivative of $x^i$. This action can be
written in the much simpler form
 \begin{equation}\label{ACTIONtwoeq}
  S = h(x(t)) - h(x(0))
 \end{equation}
which makes manifest the fact that the action is invariant under
arbitrary change in the path (provide that the endpoints are fixed). Such
a high degree of symmetry is typical of a topological theory, and is
closely related to the fact that the equations of motion, obtained by the
Euler-Lagrange process from the action, are identically satisfied. If we
turn now to the Hamiltonian formalism, we see that the conjugate momentum
to $x^i$, which we will denote by $p_i$, and define using the Euclidean
time prescription
 $p_i = i\frac{\delta{\mathcal{L}}}{\delta \dot{x}^i}$,
satisfies
 \begin{equation}\label{MOMeq}
  p_i = i \partial_i h (x), \qquad i=1, \dots,n,
 \end{equation}
where $n$ is the dimension of $\Man$,  so that there are $n$ constraints
 \begin{equation}\label{CONSTRAINTSeq}
  T_i \equiv -p_i + i \partial_i h (x) = 0, \quad i=1, \dots , n.
 \end{equation}
Using standard Poisson brackets
 $\Pb{x^i}{p_j} = \delta^i_j$, which derive from the standard symplectic form
 $\Df p_i \wedge \Df x^i$ on $T^*(\Man)$, we find that these constraints are first class,
satisfying the simple relationship $\Pb{T_i}{T_j} =0$. The number of
first class constraints is as expected equal to the number of
degrees of freedom of the field $x^i$, so that the reduced phase
space will formally have dimension zero. Proceeding now to
quantization, we use the canonical \Brst\ approach \cite{HenTei,
KosSte}, where the observables on the reduced phase space are
replaced by cohomology classes.

The canonical \Brst\ quantization of a Hamiltonian system with
classical Hamiltonian $\Hamo$ and first class constraints $T_1=0,
\dots ,T_m=0$, is derived from the observation that the observables
of the system defined by the Marsden-Weinstein reduction process
(which corresponds to Dirac brackets) are equivalent to the
cohomology classes of a function $\Brstop$ (acting by Poisson
bracket) constructed from the constraints by a standard
prescription.  This was first realised by Henneaux
\cite{Hennea,HenTei}, and then expressed in terms of Lie algebra
cohomology by Kostant and Sternberg \cite{KosSte}. To define the
\Brst\ function $\Brstop$ the original unconstrained phase space
must be extended to include anticommuting variables
$\Et^a, a=1, \dots,m$ corresponding to the constraints, known as ghosts, together
with conjugate anticommuting momentum variables $\pi_a$. Some
particular examples of this are given below. The supermanifold thus
obtained is given a super symplectic structure leading to super
Poisson brackets
 $\Pbnd{\Et^a}{\Et^b}=0, \Pbnd{\Et^a}{\pi_b}=\delta^a_b, \quad a,b=1, \dots, m$
and the \Brst\ function then defined (in the case of abelian
constraints) by
 \begin{equation}
  \Brstop = \Sum{a=1}{m} \Et^a \, T_{a},
 \end{equation}
so that the Poisson bracket
$\Pbnd{\Brstop}{\Brstop}$ is zero, and thus the cohomology of $\Brstop$  well-defined.

It is on quantization that the full power of the \Brst\ approach
emerges. The Marsden-Weinstein reduced phase space is generally very
complicated, and may not readily admit a polarisation determining
the position/momentum split. By contrast, if the original
unconstrained, symplectic manifold does have a quantization, then
the super symplectic manifold constructed for the \Brst\ formulation
also has a quantization, and the \Brst\ state space can be
identified as the space of functions of the original configuration
variables together with the ghost variables. This space of states is
graded by the degree in the ghost variables, so that ghost number
runs from $0$ to $n$, and the \Brst\ operator raises ghost number by
one.

Under this quantization, the \Brst\ function becomes an operator whose
super commutator
$\Commnd{\Brstop}{\Brstop} = \Brstop\Brstop + \Brstop\Brstop $ is zero, so that
the $\Brstop$ cohomology of states (functions) and observables
(operators) can be defined in the standard way. (The
\Brst\ cohomology of observables is the space of observables which
commute with
$\Brstop$ modulo those which are themselves commutators with
$\Brstop$.  The \Brst\ cohomology of states is defined to be the space of states
which are annihilated by $\Brstop$ modulo those which are themselves
the image of a state under the action of $\Brstop$. The action of
operators on states is then well defined at the level of
cohomology.) These constructions thus lead to a formulation in the
Schr\"odinger picture where the observables, and in particular the
Hamiltonian, are expressed as differential operators on the space of
functions on a supermanifold, and the physical quantities to be
calculated are traces over \Brst\ cohomology classes.  In order to
ensure that path integration leads to the required traces a process
known as gauge fixing is required; the general theory of
gauge-fixing is described in section \ref{GFsec}, and its
implementation for the Killing model in section \ref{GFKMsec}.
Further details of the general principals of gauge fixing may be
found in \cite{GFBFVQ}.

In the Morse theory model states are  `wave-functions'
$\psi(x,\eta)$ on the $(n,n)$-dimensional supermanifold built from
the tangent bundle of $\Man$, and the \Brst\ operator is
 \begin{equation}\label{MTBRSTeq}
   \Brstop = \eta^i T_i = i(\Df + \Et^i \partial_i h)=i e^{-h} \Df e^h,
 \end{equation}
which is the supercharge in Witten's model \cite{Witten82}.  Here we
have made the natural identification of a wave function
$\psi(x, \eta)$ with a differential form on $\Man$ obtained by identifying
$\Et^i$ with $\Df x^i$, so that $\Et^i \Pbd{x_i} = \Df$. Standard
\Brst\ techniques then suggest that the gauge-fixing fermion is
\begin{equation}
 \Gff =- i(\Da - i g^{ij} \partial_i h \pi_j) =-i e^{h} \Da e^{-h} \, .
\end{equation}
(More details of the \Brst\ construction and its
implementation are given below in sections \ref{GFsec} and
\ref{GFKMsec}.) The Hamiltonian obtained by the gauge fixing process described in
sections \ref{GFsec} and
 is the deformed Laplacian
 \begin{equation}
  \Hamg =\Textfrac{1}{2}\Comm{\Brstop}{\Gff} = \Textfrac12 (\Df+\Da)^2
    +  \Textfrac12 g^{{i}{j}} \Pbdt{h}{x^{{i}}} \Pbdt{h}{x^{{j}}}
    + \Textfrac{i}{2}g^{{i}{l}}(\eta^{{j}}{\pi_{{i}}}-{\pi_{{i}}}\eta^{{j}})
     \frac{D^2h}{Dx^{{l}}Dx^{{j}}}.
 \end{equation}

Using paths satisfying the stochastic differential equation
\begin{eqnarray}
  d\tilde{x}^{{i}}_t &=& d\tilde{b}_t -g^{{j}{i}}(\tilde{x}_t) \partial_i h(\tilde{x}_t) dt     \End
  \tilde{x}_0  &=& x
 \end{eqnarray}
together with fermionic Brownian paths leads to a simple path
integral formula which exhibits the classical action as expected,
and also leads to the appropriate Faddeev-Popov determinant in a
mathematically rigorous and direct way. Scaling $h$ by a very large
factor then leads to the topological results given by Witten in a
rigorous mathematical manner.  Details of this work may be found in
\cite{TPMT}.

Of course a trivial example of this model is obtained by setting $h$
to zero, giving as \Brst\ operator the exterior derivative $\Df$ and
as gauge fixing fermion its adjoint $\Da$ so that the gauge fixing
Hamiltonian $\Hamg$ is the Hodge-de Rham operator
$\Df \Da + \Da \Df$.  This model has been used for supersymmetric
proofs of the index theorem \cite{Alvare,FriWin}. However an
important observation of Witten is that scaling $h$ to be very large
picks out the cohomology in an interesting way, leading to the Morse
inequalities and an explicit construction of the cohomology of
$\Man$ in terms of the critical points of $h$.

Turning now to the Killing model, this involves an isometric circle
action on a Riemannian manifold $(\Man,g)$; let $X$ be the Killing vector
field which generates this circle action.

The classical field of the  theory is a map $y$ from the unit
interval into a trivial circle bundle $\Vb$ over $M$, represented in
local coordinates as
 $y(t) = (x(t),w(t))$,
where as before $x^i, i = 1, \dots, n$ are coordinates on $\Man$ while
$w$ is a coordinate on $S^1$, the fibre of $\Vb$. The radius of the
circle is chosen to be $\Conr$ so that $w$ is periodic with period
$2\pi\Conr$.

The action of the theory is
 \begin{equation}
  S((x(.),w(.)) = \int_0^1 i \Cons \dot{w}(t)  \quad \Df t
 \end{equation}
where  $\Cons$ is a real constant. This can be regarded as an
example of a generalised Morse theory model, with Morse function
$h(x,w)=i s w$ only locally defined on the fibre.

With conjugate momenta to $x^i$ and $w$ denoted respectively by
$p_i$ and $v$, there are $n+1$ first class constraints
 \begin{equation}
  T_i \equiv -p_i = 0,
  \quad U \equiv -v - \Cons =0, \quad i = 1, \dots,n.
 \end{equation}

To implement the constraints and gauge-fixing at the quantum level
we again use the \Brst\ quantization in canonical form
\cite{Hennea,HenTei}, introducing ghosts and their conjugate
momenta. For this process two supermanifolds are again required, a
super configuration space with even local coordinates
$x^i,w$ and odd local coordinates $\Et^i, \theta$ and a super phase
space with even local coordinates
$x^i,p_i,w,v$ and  odd local coordinates
$\Et^i, \pi_i, \Th,\Rh$. (In each case the index ${\scriptstyle i}$
runs from $1$ to $n$, the dimension of $\Man$, while $w$ is a
coordinate on the circle as before.) The anticommuting coordinates
$\Et^i, \Rh$ are the ghosts corresponding to the constraints
$T_i$ and $U$ respectively, while $\pi_i$ and $\Th$ are the
corresponding conjugate momenta.  The
$(n+1,n+1)$-dimensional super configuration space  is built from the
tangent bundle of
$\Vb=\Man \times S^1$, with coordinate patches
corresponding to those on $\Man$ and changes of the coordinates
$x^i,\Et^i,  \to x'{}^i,\Et'{}^i$ on overlapping coordinate patches
defined by setting $x^i \to x'{}^i(x)$ as on $\Man$ and
\begin{equation}\label{TRANSFUNCeq}
    \Et'{}^i= \Pbdt{x'{}^i}{x_j}\,\Et^j \quad  ,
  \end{equation}
while the coordinates $w$ and $\Th$ are the same in all coordinate
patches. The super phase space  is the cotangent bundle of the super
configuration space, with nonstandard coordinates such that
$x^i, \Et^i, w$  and $\Th $ transform as above,
\begin{equation}
  p'{}_i= \Pbdt{x^j}{x'{}_i}\,p_j, \quad
  \Mbox{and} \quad
  \pi'{}_i= \Pbdt{x^j}{x'{}_i}\,\pi_j \quad
   \quad ,
  \end{equation}
while the coordinates $v$ and $\Rh$ are the same in all coordinate
patches. (In each case $w$ is a coordinate on a circle of radius
$r$.)

The simplest, and natural, choice of symplectic form on this super
phase space, which makes $p_i, v, \pi_i, \Rh$ the conjugate momenta
to
$x^i,w,\Et^i,\Th$ respectively, is
 \begin{eqnarray}\label{SSYMPeq}
  \Om &=& \Df \left( p_i \wedge \Df x^i + v \wedge \Df w
  + \pi_i \wedge \Dc \Et^i + \Th \wedge \Df \Rh
  \right)
  \End
  &=& \Df p_i \wedge \Df x^i + \Df v \wedge \Df w
      + \Dc \pi_i \wedge \Dc \eta^i + \Df \Th \wedge \Df \Rh \End
   && -\frac12
    \Curv{i}{j}{k}{l}
     \eta^{k}\pi_{l}  \Df x^i \wedge \Df x^j,
 \end{eqnarray}
where the Levi-Civita connection corresponding to the Riemannian metric
$g$ has been used, with Christoffel symbols
$\Conn{i}{j}{k}$ and curvature tensor components
$\Curv{i}{j}{k}{l}$, so that
 \begin{equation}\label{CDIFeq}
  \Dc \eta^i = \Df \eta^i + \Conn{j}{k}{i} \Et^{k} \Df x^{j}
  \quad \Mbox{and}
  \quad \Dc \pi_i = \Df \pi_i - \Conn{j}{i}{k} \pi_{k} \Df x^{j} \quad  .
 \end{equation}
The corresponding Poisson brackets are:
 \begin{equation}\label{SPBeq}
 \begin{array}{lcllcllcl}
  \Pb{p_i}{x^j}& =& \delta_{i}^{j}, &
   \Pb{p_i}{p_j}   & =&  \Curv ijkl \pi_{l}\eta^{k}, &
   \Pb{\pi_{i}}{\eta^j}& =& \delta_{i}^{j},\\
 \Pb{p_i}{\eta^{j}}& =& \Conn ilj \eta^{l}, &
  \Pb{p_j}{\pi_l}& =& -\Conn jli \pi_i,  &&&\\
     \Pb{v}{w}& =& 1,  & \Pb{\Th}{\Rh} &=& 1, &&&
 \end{array}
 \end{equation}
the remainder being zero.

Quantization of this classical system is achieved by taking as wave
functions the space
$\Fun = \Omm \otimes \Vun \otimes \Thun$ where $\Vun$
is the space of smooth functions on the circle of radius $\Conr$ and
$\Thun$ is the $2$ dimensional space of functions of the single anticommuting variable $\Th$, and we have
again identified the space of functions of $x$ and
$\Et$ with the space of differential forms on $\Man$.
(Note that we are working in $\Rh$ momentum space.) The operators
$x^i, \eta^i, e^{i w/\Conr}$ and $\Th$ are simply represented by multiplication
while
 \begin{eqnarray}\label{QUANTeq}
    p_i = -i\Dc_i \equiv -i(\Pbd{x^i} + \Et^j \Conn{i}{j}{k}
    \Pbd{\Et^k}),&&
    \quad \pi_i = -i \Pbd{\eta^i} \End
   v = -i \Pbd{w} &\quad \Mbox{and} \quad &\Rh = -i \Pbd{\Th}.
 \end{eqnarray}

The \Brst\ operator
$\Brstop$ is constructed from the constraints as in (\ref{MTBRSTeq})
giving
 \begin{equation}
  \Brstop = -\eta^i p_i  - \Rh ( v + s) = i(\Df + u \Pbd{\Th})\,
 \end{equation}
where
 \begin{equation}
 u = s  -i \Pbd{w} \, .
 \end{equation}
It will be seen in the following section that this can be identified
with the differential used in a de Rham model of the equivariant
cohomology of
$\Man$ under the circle action generated by $X$.

In order to determine the correct gauge fixing fermion, it will be
useful to make the coordinate change
\begin{equation}\label{COORDCHANGEeq}
    \eta^i \to \eta^i + \Th X^i
\end{equation}
which can be implemented by conjugation by the operator
$\exp( \Th\Ix)$, where $\Ix$ denotes interior multiplication of forms along $X$,
so that the \Brst\ operator becomes
 \begin{equation}\label{DKeq}
  \Brstope
   = i (\Dke) \, ,
 \end{equation}
where $\Lx $ denotes the Lie derivative of  differential forms along
$X$. This form of the operator will be useful when determining the
appropriate choice of gauge fixing fermion.  It is closely related
to the Cartan model for the equivariant cohomology, and corresponds
to the Kalkman model \cite{Kalkma}, as will be explained below.

In order to construct the path integrals for this model, we need to
determine the correct gauge-fixing fermion.  This will be done in section
\ref{GFKMsec}, after the relationship between the \Brst\ operator and the
equivariant cohomology of $\Man$ under the circle action generated by
$X$ has been described, and general principles of gauge-fixing explained.
\section{Equivariant cohomology}\label{ECsec}
In this section, which draws heavily on the book of Guillemin and
Sternberg \cite{GuiSte}, the de Rham model of the equivariant
cohomology of a manifold under a group action is described. If a Lie
group $G$ acts on a manifold $\Man$ in such a way that the quotient
space
$\Man/G$ is a manifold, then the equivariant cohomology $\Hgm$ of $\Man$ under
the $G$ action is simply the standard cohomology of $\Man/G$. When
the $G$ action has fixed points, then $\Man/G$ will have
singularities; in such cases $\Hgm$ is defined to be the standard
cohomology of
 $(\Man \times E)/G$ where $E$ is a contractible space on which $G$ acts freely.
(It can be shown both that such a space will always exist, and that the
cohomology thus defined is independent of the choice of $E$.)

It is a standard theorem (the de Rham theorem) that the real
cohomology of $\Man$ is isomorphic to the de Rham cohomology of
$\Man$. The analogous theorem in the equivariant setting establishes
that the equivariant cohomology of $\Man$ under a $G$ action may
be realised algebraically in a similar manner, as will now be
described in the particular case where $G$ is the circle group
$\Uone$.  First, a  super Lie algebra $\Gt$ is built from the
Lie algebra of $\Uone$; if $\Gen$ is the single generator of this Lie
algebra, then the superalgebra is spanned by $\Df$, $\Gen$ and
$\Io$ with $\Gen$ even, $\Df,\Io$ odd and the only non-trivial
bracket
 \begin{equation}
   [\Df \,\Io] = \Gen \, .
 \end{equation}
Next, a further commutative superalgebra $A$ is required with
various properties. (This is the algebraic counterpart to the
space $E$.) First, there must be a representation
$\rho$ of $\Uone$ as automorphisms of $A$ and an action of $\Gt$
as super derivations of $A$, and these two actions must be compatible in
the sense that
 \begin{equation}
  \Dbd{t} \rho(\exp{t\Gen}) \bigl|_{t=0} = \Gen \, .
 \end{equation}
(In the case of the circle action on $\Man$, the algebra $\Omm$ of
differential forms on $\Man$, is such an algebra if the action of
$\Df$ on $\Omm$ is exterior differentiation, that of $\Io$  exterior
multiplication by the vector field $X$ which generates the $\Uone$ action
and that of $\Gen$ is Lie differentiation along $X$.) An additional
requirement put on $A$ is that there must exist an odd element $\Th$ of
$A$ such that  $\Io \Th = 1$. This will not be the case with $\Omm$ itself if
the $\Uone$ action has fixed points.

We now define the action of $\Df$, $\Gen$ and $\Io$ on
 $\Omm \otimes A$ by
 \begin{eqnarray}
  \Df ( \omega \otimes a) &=& \Df  \omega \otimes a +   \omega \otimes \Df a \End
  \Gen ( \omega \otimes a) &=& \Lx  \omega \otimes a +   \omega \otimes \Gen a \End
  \Io ( \omega \otimes a) &=& \Ix  \omega \otimes a +   \omega \otimes \Io a
 \end{eqnarray}
and define an element $f$ of $\Omm \otimes A$ to be {\sl basic} if
both $\Io f = 0$ and $\Gen f = 0$. The operator $\Df$ acts on
$\Bas$, the set of such elements, and, provided that $A$ is acyclic, we have
the equivariant de Rham theorem
 \begin{The}
 $\Hgm = H(\Bas, \Df)$.
 \end{The}
In the proof of this theorem given by Guillemin and Sternberg
\cite{GuiSte} it is shown that the cohomology
$H(\Bas, \Df)$ is independent of the choice of $A$ satisfying the
various conditions.  The acyclicity of $A$ is the algebraic analogue of
the contractibility of $E$ in the geometric definition of equivariant
cohomology given above, while the existence of
$\Th$ such that $\Io \Th = 1$ is the analogue of the free action of
$G$ on $E$.

A particular choice of the super algebra
$A$ is the Weil algebra $\Wm$ defined below; the corresponding
model of the equivariant cohomology will be shown to be the
topological quantum theory constructed in the previous section, with
the modified version of the theory achieved by change of coordinate
corresponding to the Kalkman model, also described below.

 \begin{Def}
The Weil algebra $\Wm$ is the $\mathbb{Z}$-graded superalgebra
 \begin{equation}
  \Wm = \Lambda(\Guts)\otimes S(\Guts)
 \end{equation}
where $\Lambda$ denotes the exterior algebra and $S$ the symmetric
algebra.  The action of
$\Uone$ on this algebra is defined to be the coadjoint action.
Elements of this space will be written as formal sums
$\sum_{l=0}^{\infty} (a_l + \Th b_l)u^l$ where, if $x$ is the single element of a basis of $\Guts$,
$u=1 \otimes x$ and $\Th=x \otimes 1$. The action of the elements of $\Gt$ on $\Wm$ is defined by
 \begin{equation}
   \Io_W =  \Pbd{\Th}, \quad
   \Df_W = u \Pbd{\Th}, \quad \Mbox{and} \quad \Gen_W = 0 \, .
 \end{equation}
 \end{Def}
It is almost immediate that  $\Wm$ is acyclic and
$\Io_W \Th = 1$, so that a model for the equivariant cohomology
of
$\Man$ under the given $\Uone$ action is the cohomology of $\Basw$
under the operator
 \begin{equation}
  \Dw = \Df + u \Pbd{\Th} \, .
 \end{equation}
Explicit calculation (as in \cite{AtiBot}) shows that the space
$\Basw$ consists of elements of the form
 $$
 \sum_{l=0}^\infty (\omega_{l} - \Th\, \Ix\, \omega_{l})u^l
 $$
where  each $\omega_{l}$ is a form in $\Omm$ such that
$\Lx \omega_l = 0$.

A useful variation of this model is  the Kalkman  model
\cite{Kalkma}. This is defined by setting
$\lambda:\Omm \otimes \Wm \to \Omm \otimes \Wm$ with
$\lambda=e^{\Th \, \Ix}$ so that $\Dw$ is replaced by the Kalkman differential
 \begin{equation}
 \Dk = \Dke \,
 \end{equation}
while the basic algebra is simply the subset $K$ of
$\Omm \otimes \Wm$ consisting of the $\Th$ independent $\Uone$ invariant elements.

This construction leads directly to the  Cartan model of the
equivariant cohomology; this is the cohomology of
$\Uone$ invariant elements of the algebra
 $\Omm \otimes S(\Guts)$ with differential
 $\Dcc = \Df + u \Ix$ \cite{Kalkma,AtiBot,GuiSte}, where $u$ is the single generator of
$\Guts$, the dual of the Lie algebra of $\Uone$. However it will
be useful to use the Kalkman model since this still has a usefully
simple criteria for a form to be basic, but also retains a
differential whose square is zero on the full algebra
$\Omm \otimes \Wm $ and not simply on the subalgebra from which the equivariant
cohomology is built.

This can now be related directly to the \Brst\ cohomology of the
Killing model, in the form introduced at the end of section
\ref{MODsec}, in which the
\Brst\ operator was shown to take the form
\begin{equation}
  \Brstope
  = i(\Dke) \,
 \end{equation}
that is, it is the Kalkman differential acting on the space
$\Omm \otimes\Gun$, which
is isomorphic to $\Omm \otimes \Wm$, with the formal generators
$u,\Th$ of the Weil algebra identified with the operators $u,\Th$ in the topological model.

Of course the physical model does not have the restriction to the
basic subalgebra built into it; this is achieved by careful choice
of gauge fixing, as will be described in the following sections. The
version of the model we have constructed allows a simple
gauge-fixing term to lead to correct path integral calculations.
\section{Gauge fixing, general principles}\label{GFsec}
In order to implement \Brst\ quantization by path integral methods
a mechanism known as gauge fixing is required to ensure that the
traces calculated by the path integrals are traces over \Brst\
cohomology classes, and thus (as explained in section
\ref{MODsec}) over the physical states of the model. The gauge
fixing mechanism involves adding a term $\Hamg$ to the Hamiltonian
which, while zero in the operator cohomology, performs the
analytic function of ensuring that all necessary operators are
trace class so that the cancellations which formally ought to
occur because of super symmetry actually do occur.  This
gauge-fixing term $\Hamg$ in the Hamiltonian is the super
commutator
$\Commnd{\Brstop}{\Gff}= \Brstop\Gff+ \Gff\Brstop$ of an odd operator $\Gff$ known as the
gauge-fixing fermion with the \Brst\ operator $\Brstop$. An
extended account of gauge fixing in canonical \Brst\ quantization
may be found in \cite{GFBFVQ}.

The first step in establishing this mechanism is to observe that
if $A$ is an even observable for the \Brst\ quantum system, so
that
$\Comm{A}{{\Brstop}}=0$, then, as first observed by Schwarz, \cite{Schwar},
the supertrace of $A$ over all states is equal to the supertrace of
$A$ over the \Brst\ cohomology classes of states. If
$H^i(\Brstop)$ denotes the cohomology of $\Brstop$ at ghost number $i$,
this result may be expressed by the equation
 \begin{equation}\label{LTeq}
  \Strace\,\, A = \sum_{i=0}^{n} (-1)^i \Trace_{H^i(\Brstop)} A,
 \end{equation}
where $\Strace\,\, A $, the supertrace of the operator $A$, is the
trace of
$\epsilon A$, where $\epsilon$ is the grading operator
which has eigenvalue $+1$ on  states of even ghost number and $-1$ on odd
states. A formal proof comes from noticing that there is cancellation in
the supertrace between eigenvalues of $A$ corresponding to states which
are not in the kernel of $\Brstop$ and eigenvalues of states which are in
the image of
$\Brstop$.  This follows from the observation that if $A f = \lambda f$
then
$A \Brstop f = \lambda \Brstop f$.  Thus the only eigenvalues
which survive to contribute to the supertrace are those
corresponding to
\Brst\ cohomology classes.

This naive argument may of course break down if the reordering of
the infinite sums in the traces is not valid.  However,  by
combining $A$ with the operator
$\exp-\Hamg \, t, \, t>0$ with $\Hamg=\Comm{\Brstop}{\Gff}$
constructed from an appropriate gauge fixing fermion $\Gff$ it is
possible to obtain precisely the desired traces over cohomology
classes, provided  of course that these traces exist. (In this paper
all cohomology classes will be finite-dimensional so this issue does
not arise.) The key property which the gauge-fixing Hamiltonian
$\Hamg$ must possess  is that the operator
$\exp-\Hamg \, t$ is trace class for all positive $t$. With this condition satisfied
the alternating sums in (\ref{LTeq}) are all absolutely
convergent, and the reordering used to establish this equation is
valid.  Further details of these ideas may be found in
\cite{GFBFVQ}.

A simple example, familiar from differential geometry, of a \Brst\
operator and gauge-fixing fermion is the exterior derivative $\Df$
on the space of forms on a manifold, and its adjoint $\Da$.  The
gauge-fixing Hamiltonian $\Hamg$ is then the Hodge de Rham operator
$\Comm{\Df}{\Da}$, the \Brst\ cohomology is the de Rham cohomology,
and Hodge theory ensures that $\Hamg$ has the necessary
properties.  The relationship between the supertrace and de Rham
cohomology was first pointed out in this context by McKean and
Singer \cite{MckSin}. The Morse theory model described in
\cite{Witten82} has a closely related
\Brst\ operator $\Brstop = i e^h \Df e^{-h}$, and gauge fixing fermion
$\Gff=-i e^{-h} \Da e^h$.
\section{Gauge fixing in the Killing model}\label{GFKMsec}
In this section we construct the gauge fixing fermion which ensures
that path integral calculations for observables in the Killing model
reduce correctly to traces over the equivariant cohomology of
$\Man$ under the circle action generated by $X$.

As already established in section \ref{ECsec}, this equivariant
cohomology can be expressed as the cohomology of the
\Brst\ operator for the Killing model, which takes the form
(equation (\ref{DKeq}))
 \[
   \Brstope
  = i(\Dke)\, ,
 \]
acting on the space square integrable functions on the super
configuration space  which are independent of $\Th$ and are zero
eigenstates of $\Lx$.  As gauge-fixing fermion we take
 \begin{equation}\label{GFeq}
   \Gffe =- i\left( \left(\Da + \Ua \Ix\Adjd \right)(1 - \Th \Pbd{\Th})
    + {\Conlp} (\Ua-s) \Th - \Conl \Lxa \Pbd{\Th}\right)\, .
 \end{equation}
This choice of gauge-fixing fermion will be justified below. The
gauge-fixing Hamiltonian is
 \begin{eqnarray}\label{HAMeq}
   \Hamg = && \Comm{\Brstope}{\Gffe}  \End
         = && \left(\Comm{\Df}{\Da} +  u \Comm{\Ix}{\Da}
         +  \Ua \Comm{\Ix\Adjd}{\Df}  +   u^2 \Comm{ \Ix}{  \Ix\Adjd}\right)
         (1 - \Th \Pbd{\Th})
            \End
         && \quad +    (\Da + \Ua \Ix\Adjd)\Pbd{\Th}
         +\Lx(\Da + \Ua \Ix\Adjd) \Th
         + {\Conlp}u (\Ua-s) + \Conl \Lx\Lxa \, .
 \end{eqnarray}
The effect of the gauge-fixing fermion and the corresponding
Hamiltonian can be seen by recalling that the space on which this
operator acts is
$\Fun =\Omm \otimes \Vun \otimes \Thun$ where
$\Vun$ is the space of smooth functions on $S^1$ and $\Thun$ is the
space of functions of the single anticommuting variable $\Th$. The
operator
$1-\Th\Pbd{\Th}$ acts on $\Thun$ as a projection onto the one dimensional subspace
of functions independent of $\Th$. On $\Vun$ the operators $u$ has
eigenvalues  $s+ 2\pi k/\Conr$ with $k$ an integer. The role of the
$u(\Ua-s)$ term is to project onto the
$s$ eigenspace of $u$  in the supertrace,
while the $\Lx\Lxa$ term projects onto the $\Uone$ invariant forms
in $\Omm$. On any
$u$ eigenspace the operator
 $$H_s =\Comm{\Df}{\Da} +  u \Comm{\Ix}{\Da}
         +  \Ua \Comm{\Ix\Adjd}{\Df}  +   u^2\Comm{ \Ix}{  \Ix\Adjd} =
         \Comm{\Df+u\Ix}{\Df\Adjd + \Ua\Ix\Adjd}$$
         has non-negative
eigenvalues. Relative to the basis of $\Thun$ consisting of the
function $1$ and the function $\Th$, both the term
$i(\Da + \Ua \Ix\Adjd)\Pbd{\Th}$ and the term
$-i\Lx(\Da + \Ua \Ix\Adjd) \Th$ are off diagonal, and their product is
$u^2\Lx\Lxa$.  Provided that $\Cons<\pi/2\Conr$ the greatest
negative contribution from these off diagonal terms cannot outweigh
the positive contribution from the terms
${\Conlp}u (\Ua-s) $ and $ \Conl \Lx\Lxa$ so that $\Hamg$ can only
have a zero eigenvalue on a state which is an eigenstate of $u$ with
eigenvalue $s$ and is a zero eigenstate of $\Lx$.

Thus we have shown that the supertrace of $\exp(-\Hamg t)$ will
reduce to a supertrace over the cohomology of the basic subspace of
$\Omm \otimes \Wm$, with the formal variable $u$ replaced by the
non-zero real number $\Cons$. In other words, we have obtained a
supertrace over the equivariant cohomology of $\Man$ under the
$\Uone$ action generated by $X$. We have thus constructed the
desired gauge-fixing Hamiltonian, and have the theory in a form
which allows calculations by the superspace path integral methods
developed in \cite{SCSTWO,SBMS}. As further confirmation,  after
taking the supertrace in the
$u$ and $\Th$ sector we obtain
\begin{equation}
    \Hamg = \Comm{\Df+s\Ix}{\Df\Adjd + s\Ix\Adjd}
\end{equation}
as in the Witten model \cite{Witten82}.  Path integral
quantization can be carried out rigorously using the methods of
\cite{SCSTWO,SBMS}.
\section{Morse theory on the odd cotangent bundle}\label{MTOCBsec}
An alternative derivation of the Killing model, which brings in the
$\Uone$ action from the outset,
is to use the moment map of the $\Uone$ action on the odd cotangent
bundle
$\Odd{T^*\Man} $
of
$\Man$.  The odd cotangent bundle is the $(n,n)$-dimensional
supermanifold built from the cotangent bundle of $\Man$. It has
local coordinates $x^i$ as on $\Man$ and $\pi_i$ transforming as
the components of a one form on $\Man$.  The moment map
$p: \Odd{T^*\Man} \to \Guts$ with respect to the odd symplectic structure
$\tilde{\omega} = \Df\pi_i \wedge \Df x^i$ can then be expressed as
$X^i(x) \pi_i$.

If we now consider the $(n,n+1)$-dimensional supermanifold
$ \Odd{T^*\Man} \times \Real^{0,1}$, parametrising $\Real^{0,1}$ by a
single odd variable
$\Th$, and  construct the Morse theory model as in (\ref{ACTIONeq}) for the function
$\Th X^i(x) \pi_i$, we obtain the action
\begin{eqnarray}
  \lefteqn{S(x(.), \pi(.),\Th(.))} \End
  &=& \int_0^t \dot{\Th}(t') X^i(x(t')) \pi_i(t')
  + \Th(t') \partial_j X^i(x(t')) \dot{x}^j(t') \pi_i(t')\End
 && \quad\quad + \Th(t') X^i(x(t')) \dot{\pi}_i(t')\, \Df t' \, .
\end{eqnarray}
This model has abelian first class constraints
\begin{eqnarray}
 T_i \equiv && -p_i +i \partial_i X^j(x(t'))  \pi_j(t') \End
 \tau^i \equiv  && -\Et^i -i \Th(t') X^i(x(t')) \End
 \Mbox{and} \quad
 \upsilon \equiv && -\Rh +i X^j(x(t')) \pi_j(t')
\end{eqnarray}
where $p_i, \Et^i$ and $\rho$ are the canonically conjugate
momenta to $x^i, \pi_i$ and $\Th$ respectively; on quantization we
obtain as
\Brst\ operator for the model
 \begin{eqnarray}
 {\Brstop}
  =&& e^{-\Th X^i \pi_i} (-\alpha^i p_i - q_i \Et^i - u \Rh ) e^{\Th X^i \pi_i}
 \end{eqnarray}
where $\alpha^i, q_i$ and $u$ are the ghosts for the constraints
$T_i, \tau^i$ and $\upsilon$ respectively.
If we make the identification $\alpha^i=\Texthalf\Et^i$ and
$q_i = \Texthalf p_i$ we obtain
 \begin{eqnarray}
 {\Brstop}
 = && i e^{-\Th X^i\pi_i} (\Df + u \Pbd{\Th}) e^{\Th X^i\pi_i}
  \end{eqnarray}
which we recognise as the \Brst\ operator for the Killing model
expressed in the Kalkman form (\ref{DKeq}).  An outstanding issue
is to justify the identifications made above; it seems likely, in
view of the odd symplectic form $\tilde{\Om}$ and the presence of
extra fields $\theta$ of ghost number $1$ and $u$ of ghost number
$2$, that this will involve the
\Bv\ procedure \cite{BatVil81,Nerses98}.
%
 
%
 \end{document}